\documentclass{sig-alternate-2013}

\permission{\copyright 2017 International World Wide Web Conference Committee \\ (IW3C2), published under Creative Commons CC BY 4.0 License.}
\conferenceinfo{WWW'17 Companion,}{April 3--7, 2017, Perth, Australia.}
\copyrightetc{ACM \the\acmcopyr}
\crdata{978-1-4503-4914-7/17/04. \\
http://dx.doi.org/10.1145/3041021.3054221 \\
\includegraphics{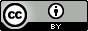}}

\usepackage{subfigure}
\usepackage{algorithm}
\usepackage{algorithmic}
\usepackage{natbib}
\usepackage{dsfont}
\usepackage{epstopdf}
\usepackage[usenames,dvipsnames]{xcolor}
\usepackage[bookmarks=false]{hyperref}
\hypersetup{
  pdftex,
  pdffitwindow=true,
  pdfstartview={FitH},
  pdfnewwindow=true,
  colorlinks,
  linktocpage=true,
  linkcolor=Green,
  urlcolor=Green,
  citecolor=Green
}
\usepackage{times}

\usepackage[backgroundcolor=White,textwidth=0.5in]{todonotes}

\usepackage[capitalize]{cleveref}

\newcommand{\commentout}[1]{}
\newcommand{\junk}[1]{}

\renewcommand{\qed}{\rule{5pt}{5pt}}

\newcommand{\realset}{\mathbb{R}}

\newcommand{\E}[2]{\mathbb{E}_{#2} \! \left[#1\right]}

\newcommand{\I}[1]{\mathds{1} \! \left\{#1\right\}}

\newcommand{\set}[1]{\left\{#1\right\}}

\mathchardef\mhyphen="2D

\usepackage{chngpage}

\begin{document}






%

\title{Does Weather Matter? Causal Analysis of TV Logs\vspace{-0.1in}}


\numberofauthors{1}
\author{
%
%
\alignauthor
Shi Zong$^\dag$\quad Branislav Kveton$^\ddag$\quad Shlomo Berkovsky$^\S$\quad Azin Ashkan$^\sharp$\quad Nikos Vlassis$^\ddag$\quad Zheng Wen$^\ddag$ \\
       \affaddr{$^\dag$The Ohio State University, Columbus, OH, USA\quad 
       $^\ddag$Adobe Research, San Jose, CA, USA}\\
       \affaddr{$^\S$CSIRO, Epping, NSW, Australia\quad 
       $^\sharp$Google Inc., Mountain View, CA, USA}\\
       \email{ zong.56@osu.edu, \{kveton, vlassis, zwen\}@adobe.com, Shlomo.Berkovsky@data61.csiro.au, azin@google.com}
}

\maketitle

\vspace{-1in}

\begin{abstract}
Weather affects our mood and behaviors, and many aspects of our life. When it is sunny, most people become happier; but when it rains, some people get depressed. Despite this evidence and the abundance of data, weather has mostly been overlooked in the machine learning and data science research. This work presents a causal analysis of how weather affects TV watching patterns. We show that some weather attributes, such as pressure and precipitation, cause major changes in TV watching patterns. To the best of our knowledge, this is the first large-scale causal study of the impact of weather on TV watching patterns.
\end{abstract}

%
%

%
%

%
%



\section{Introduction}
\label{sec:intro}

Weather affects our mood, and thus human behaviors. One of the pronounced examples is the \textit{seasonal affective disorder} -- prolonged lack of sunlight that can depress people \cite{partonen1998seasonal}. Weather indirectly affects various aspects of our lives: work and study, purchasing behaviors, and more. In this work, we set out to examine the effects of weather on TV watching patterns. In particular, we explore whether people watch \textit{different genres} of programs in different weather conditions.

This knowledge can have several important implications. First, marketers may be willing to adjust the content and ratio of the advertisements to the target audience it will be exposed to \cite{farahat2012effective}. 
Second, TV and video recommender systems may leverage this knowledge and adapt their recommendations accordingly \cite{XuBATMK13}.

To the best of our knowledge, our work is the first to apply causal analysis to a nation-scale dataset of TV consumption logs containing about $10$M watching events of more than $40$k users. We propose several practical ways for estimating the causality of weather on TV watching behavior, and observe high correspondence between our findings.


\section{Causal Analysis}
\label{sec:methods}

The problem of estimating causal effects from observational data is central to numerous disciplines \cite{pearl2009causality}. It can be formalized as follows. Let $\set{1, \dots, n}$ be a set of $n$ units $i$, such as individuals. Let $T_i \in \set{0, 1}$ indicate the treatment of unit $i$. That is, $T_i = 0$ if unit $i$ is \emph{control} and $T_i = 1$ if the unit is \emph{treated}. Then unit $i$ has two \emph{potential outcomes}, $Y_i(1)$ if the unit is treated and $Y_i(0)$ otherwise. The unit-level causal effect of the treatment is the difference in potential outcomes, $\tau_i = Y_i(1) - Y_i(0)$, and the \emph{average treatment effect on treated (ATT)} is $\E{\tau_i}{i: T_i = 1} = \E{Y_i(1)}{i: T_i = 1} - \E{Y_i(0)}{i: T_i = 1}$. Note that $\E{\tau_i}{i: T_i = 1}$ cannot be directly computed, because $Y_i(0)$ is unobserved in treated units $\set{i: T_i = 1}$.

Since the assignment to treatment and control groups is usually not random, $\E{Y_i(0)}{i: T_i = 0}$ is a poor estimate of $\E{Y_i(0)}{i: T_i = 1}$. A key challenge in causal analysis is to eliminate the resulting imbalance between the distributions of treated and control units. A popular approach to balancing the two distributions is \emph{nearest-neighbor matching (NNM)} \cite{rubin1973matching}. In this work, we match each treated unit to its nearest control unit based on their covariates, and then the response of the matched unit serves as a \emph{counterfactual} for the treated unit. In particular, the ATT is estimated as
\begin{equation}
  \textstyle
  \text{ATT} \approx \frac{1}{n_T} \sum_{i: T_i=1} (Y_i(1) - Y_{\pi(i)}(0)),
  \label{eq:treated_att}
\end{equation}
where $n_T = \sum_{i = 1}^n T_i$ is the number of treated units, $Y_i(1)$ is the observed response of treated unit $i$, and $Y_{\pi(i)}(0)$ is the observed response of the \emph{matched} control unit $\pi(i)$. The covariate of unit $i$, ${\bf x}_i$, should be chosen such that its potential outcomes are statistically independent of $T_i$. In this case, the estimate in \eqref{eq:treated_att} resembles that of a randomized experiment.


\section{Experiments}
\label{sec:exp}

In this work, we use a dataset gathered by an Australian national IPTV provider. We obtained the complete Australia-wide logs for a period of $26$ weeks, from February to September 2012 \cite{XuBATMK13}. The dataset contains 
$10$M viewing events of about 
$40$k users, who watched more than $11$k unique programs in $14$ TV genres. We also obtained the geographic locations of the users and matched them with their weather. We further extracted eight \emph{attributes} that characterize various aspects of weather (\cref{tb:treat}).

\begin{figure*}[t]
  \centering
  \subfigure[Normalized ATTs of pressure.]{\label{fig:att_pressure}
  \includegraphics[width=0.27\textwidth]{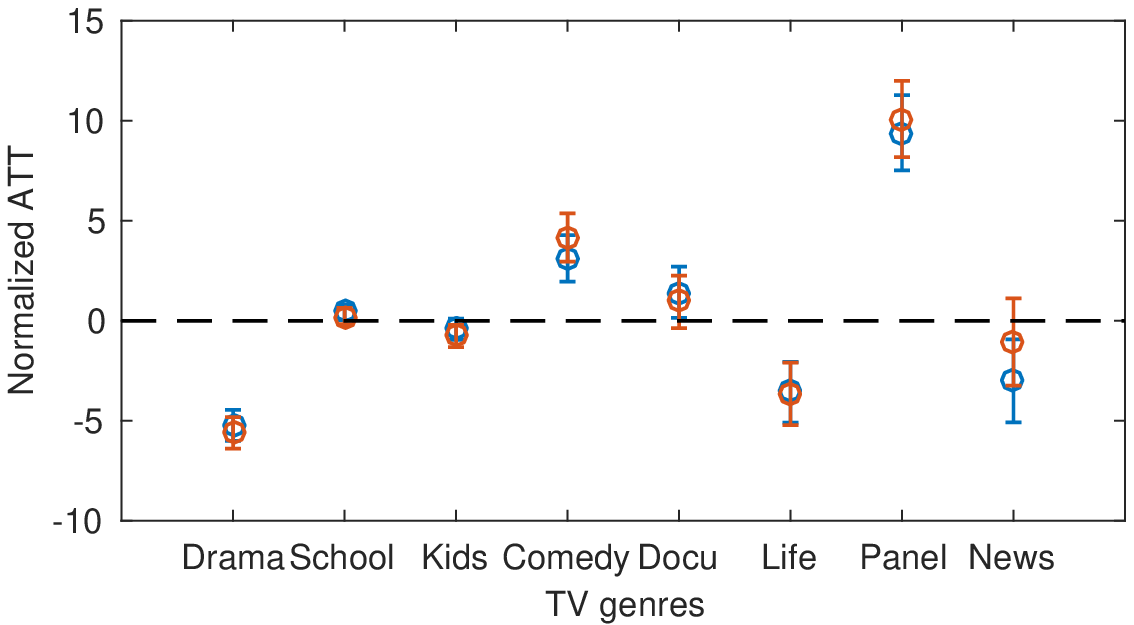}
  }
  \,
  \subfigure[Statistically significant changes in normalized ATTs for all weather-genre combinations.]{\label{fig:att_all}
  \includegraphics[width=0.35\textwidth]{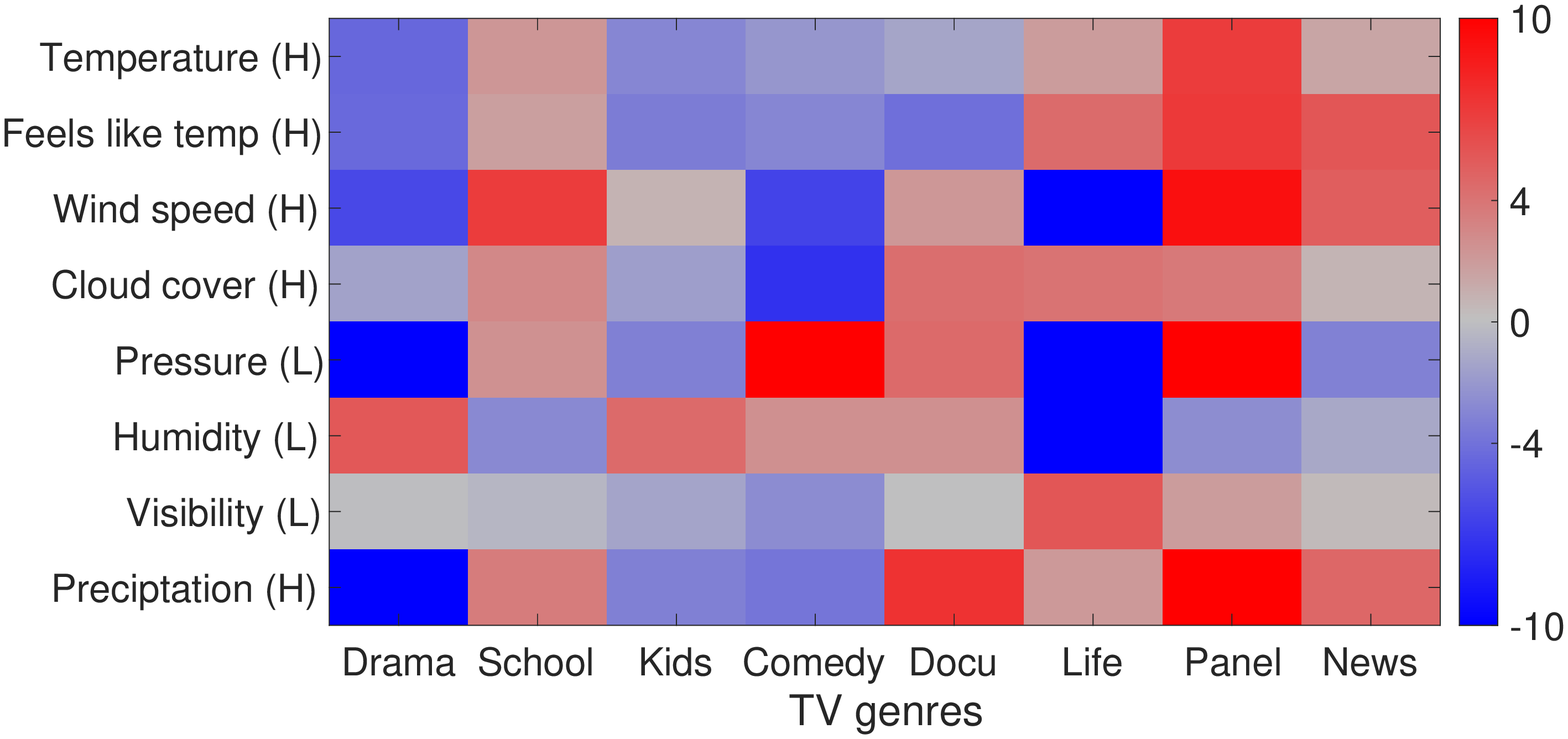}
  \hspace{-0.1in}
  \includegraphics[width=0.35\textwidth]{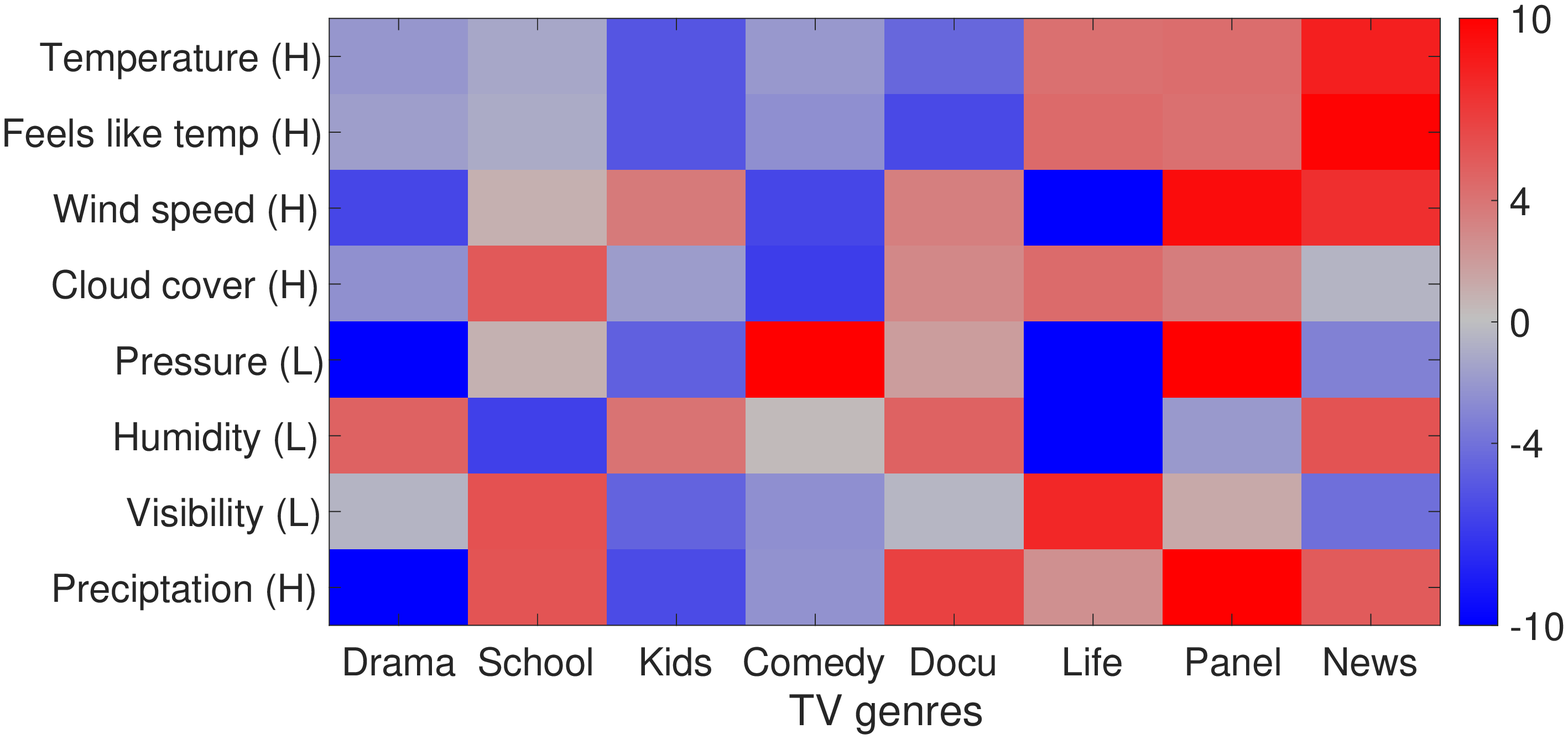}}
  \vspace{-0.1in}
  \caption{(a) Normalized ATTs of pressure on $8$ most popular genres with TV-genre (blue) and latent (red) user profiles. (b) Significant changes in normalized ATTs of all weather treatments on $8$ most popular genres. Significant increases are red, significant decreases are blue, and insignificant effects are gray. We report results for both TV genre (left) and latent (right) user profiles.}
\end{figure*}

\subsection{Experimental Setup}
\label{sec:setup}

We apply the causality framework from \cref{sec:methods} to our dataset. As we explain our setup, we illustrate it on an example query ``does high temperature cause watching more Dramas?''. The \emph{units} $i$ are TV watching events and we estimate the causal effect of weather at these events. The \emph{treatment} $T_i$ indicates the treatment weather at event $i$. In our example, $T_i = \I{\text{high temperature at event $i$}}$. We define one treatment variable for each weather attribute, and get eight weather-attribute treatments. For each treatment, we treat the events in the tail of the distribution of that attribute. That is, if the tail of the distribution is on the left (right), we assign $20\%$ of the events with the lowest (highest) values to the treatment group. We list our treatment groups in \cref{tb:treat}. The value of $20\%$ is chosen so that the number of treated events is sufficiently large.

The \emph{potential outcomes} at event $i$ under control and treatment, $Y_i(0)$ and $Y_i(1)$, are the indicators of watched TV content under control and treatment, respectively. In our example,
\begin{align*}
  Y_i(0) & = \I{\text{Drama watched at event $i$ if the temperature is low}}, \\
  Y_i(1) & = \I{\text{Drama watched at event $i$ if the temperature is high}}.
\end{align*}
Then the ATT in \eqref{eq:treated_att} is the change in the frequency of watching Dramas due to temperature. If the ATT is significantly above (below) zero, we say that high temperature increases (decreases) the frequency of watching Dramas. A near-zero ATT means that there is no effect on Dramas. We define potential outcomes and ATTs for all combinations of TV genres and treatments. 
Because some genres are infrequent, we report ATTs divided by the frequency of their genres. This \emph{normalization} is only for visualization purposes and has no impact on the statistical significance of our findings.

\begin{table}[t]
  \centering
  {\small
  \begin{tabular}{l|c||l|c} \hline
    Weather attribute & Treated &  Weather attribute & Treated\\ \hline
    Temperature & High (H) & Pressure & Low (L) \\
    Feels-like temperature & High (H) & Humidity & Low (L) \\
    Wind speed & High (H) &  Visibility & Low (L) \\
    Cloud cover & High (H) & Precipitation & High (H)\\\hline
  \end{tabular}
  }
  \caption{Our weather attributes with their treatment groups.}
  \label{tb:treat}
  \vspace{-0.15in}
\end{table}

The covariate ${\bf x}_i$ is the profile of the user at event $i$ and the time of the event. Both strongly affect $(Y_i(0), Y_i(1))$, and therefore are good candidates for guaranteeing that $(Y_i(0), Y_i(1))$ and $T_i$ are independent. We experiment with two kinds of user profiles. The first is a vector of the frequencies of watched TV genres. This profile captures high-level genre preferences. The second profile is estimated from matrix $M \in \set{0, 1}^{n_u \times n_p}$ that indicates which programs are watched by which users, where $n_u$ is the number of users and $n_p$ is the number of programs. Let $U \in \realset^{n_u \times d}$ be the user later factors in rank-$d$ truncated SVD of $M$. Then the profile of user $u$ is the $u$-th row of $U$. We choose $d = 16$. The eigenvalues of $M$ are small afterwards, and hence the first $16$ singular vectors of $M$ capture most of its structure.

\subsection{Results}

\cref{fig:att_pressure} shows the normalized ATTs \eqref{eq:treated_att} of pressure on $8$ popular genres. We observe that when the pressure is low, the frequency of watching Dramas decreases by $5\%$. The same effect is observed for both user profiles, which strengthens the validity of our results.

\cref{fig:att_all} show the significant effects for all weather-genre combinations. We measure the significance of the ATT by its estimate divided by its standard error, which is its standard deviation. Therefore, the value of $4$ ($-4$) means that the estimate is $4$ standard deviation above (below) zero, and significant at $p \approx 10^{-4}$. We observe that many significant changes in \cref{fig:att_all} are stable across our two profile representations. The correlation coefficient between the values in the two plots is as high as $0.949$, which indicates that the causal dependencies are aligned regardless of the profile representation. This reaffirms the validity of our results.

\cref{fig:att_all} reveals some insightful trends. For example, when the feels-like temperature is high, we observe an increase in News and a decrease in Kids programs. One plausible explanation for this is that warm days are suitable for outdoor activities and kids watch less TV. Therefore, the volume of the genres watched by adults increases. When the precipitation is high, we observe a decrease in Dramas and an increase in Panels. One plausible explanation for this is that rainy days make people less happy, and they watch less Dramas that are unlikely to cheer them up. 

We also validate the quality of our matching. The average Euclidean distance between the covariates of randomly matched events is $0.72$. The distance of our matched pairs is $0.13$. This $82\%$ improvement indicates that our matching is very good, and that the estimated effects may be causal.





\section{Conclusions}
\label{sec:conclusions}

In this work, we conduct a causal analysis of weather on TV watching patterns. We use two user profile representations that reveal similar dependencies. To the best of our knowledge, this is the first causal analysis of a large-scale data that looks at the interplay between weather and watching TV.

It is worth noting that our work is domain-agnostic. However, some attributes, such as humidity and precipitation, are clearly correlated. We posit that some domain knowledge could help us to refine the set of influential attributes and substantially improve our results. We believe that the uncovered weather dependencies can be incorporated into a context-aware recommender to enhance its recommendations, and we leave this for future work.


\bibliographystyle{abbrv}
\bibliography{References}

\clearpage

\end{document}